# End to End Verification and Validation with SPIN

Asankhaya Sharma

Department of Computer Science
National University of Singapore
asankhs@comp.nus.edu.sg

*Abstract* — **Over the last several years the tools used for model checking have become more efficient and usable. This has enabled users to apply model checking to industrial-scale problems, however the task of validating the implementation of the model is usually much harder. In this paper we present an approach to do end to end verification and validation of a real time system using the SPIN model checker. Taking the example of the cardiac pacemaker system proposed in the SQRL Pacemaker Formal Methods Challenge we demonstrate our framework by building a formal model for the cardiac pacemaker in SPIN, checking for desirable temporal properties of the model (expressed as LTL formulas), generating C code from the model (by refinement of PROMELA) and validating the generated implementation (using SPIN). We argue that a state of the art model checking tool like SPIN can be used to do formal specification as well as validation of the implementation. To evaluate our approach we show that our pacemaker model is expressive enough to derive consistent operating modes and that the refinement rules preserve LTL properties.**

*Keywords- Model Checking, Verification, Validation, Pacemaker*

## I. INTRODUCTION

Formal methods have been used to validate requirements and designs of software systems typically early in the development lifecycle. The use of formal methods for software and hardware design is motivated by the expectation that, as in the other engineering disciplines, performing appropriate mathematical analysis can contribute to the reliability and robustness of a design [8]. However, the high cost of using formal methods means that they are usually only used in the development of high-integrity systems [9] where safety or security is of utmost importance. Over the last several years many tools have been built that aid in the formal modeling and model checking of software. One of the most widely used model checkers is the SPIN tool. SPIN is a state of the art model checker, several years in development and has been used to do modeling of a wide variety of real time systems.

In this paper we show how a modern model checking tool like SPIN can be used to not only formally specify and model a real time system but also to aid in the generation of an implementation as well as validation of the code. For illustration of our method we choose the cardiac pacemaker system [1], for which the informal specification has been released in the public domain. Throughout the paper we present our approach as applied to the cardiac pacemaker system, in particular our main contributions are

—Modeling the pacemaker system in PROMELA
—Verification of desired properties expressed in LTL
—Generation of C code by refinement of the PROMELA model
—Validation of the implementation against the same set of desired properties

As a consequence of the above we show that it is possible to do an end to end verification of the system by using only a state of the art model checker (SPIN). The key idea behind our approach is that several years of development on formal method tools have added sufficient functionality and usability that now we can use the tools not only for model checking but also for complete verification and validation. The rest of the paper describes the verification of the cardiac pacemaker system and is organized as follows; in section II we describe the pacemaker system and its components, in section III we describe our PROMELA model, in section IV we show the verification results for a set of desired properties for the model, section V demonstrates how to generate code from the model and to validate it, in section VI we present the evaluation of our approach and results, section VII discusses some related work and finally we conclude in section VIII.

## II. THE PACEMAKER SYSTEM

The pacemaker challenge was issued by SQRL, McMaster University in 2007 [15]. An artificial cardiac pacemaker is a critical system which is used to treat patients with various heart conditions in which the natural pace generation is affected. The device is actually implanted inside the patient's body and generates stimulated paces to the heart using electric impulses. As in any critical system it is of utmost importance that the device itself should not contain any bugs or defects. A good way to ensure the reliability of the device is to use formal methods in order to capture and validate important requirements for the functioning of the device. Most of the devices are proprietary and the implementation details are only known to the manufacturer. In order to facilitate the use of formal methods and to encourage participation from a wide audience Boston Scientific released the system specification into public domain, [1] for a previous generation pacemaker. This informal specification is the basis of the modeling and verification results described in this paper. A pacemaker system consists of several components. Some of them are illustrated in Fig. 1.

Among the various components (in Fig. 1) the most critical ones are on the left. These components are physically implanted inside a patient via surgery. The focus of our approach is to use formal methods to model and validate these components since they are most important for the entire pacemaker system to function. Another aspect of this problem is to model the behavior of the heart (or environment) which interacts with the pacemaker in different ways. Some of the heart conditions are characterized by the kind of problems in the heart and others capture the normal functioning of a human heart.

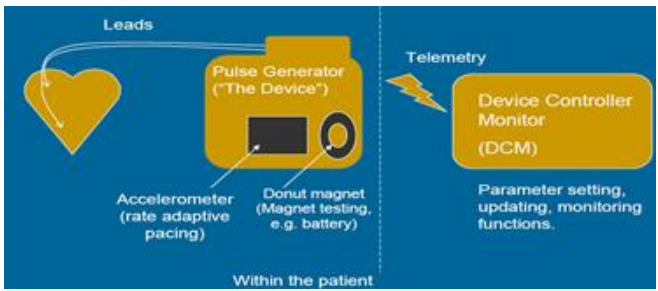

Figure 1. Components of a pacemaker [1]

Timing requirements play a major role in pacemaker software, since the pulses to heart have to be delivered at specified intervals and failure to do so may be fatal. There are several programmable parameters in the pacemaker system which provides flexibility to control the rate, time, amplitude and other properties of a pulse. We have tried to capture all the parameters which may influence the timing requirements of the system.

### III. PACEMAKER PROMELA MODEL

We follow the incremental development guidelines for building pacemaker models as suggested in [3]. In the SPIN modeling tool each component of the pacemaker can be modeled as a different process. These processes are specified in PROMELA, which captures the behavior of the processes. In order to communicate with each other these processes can use global variables or channels. Our report [13] describes all the models (sequential, concurrent and distributed) in detail, in this paper we use the model shown in Fig. 2. In this model each process can read and write any of the global variables illustrated by the double sided arrows. The Heart process captures the environment which is in constant interaction with the Pace Generator via a Sensor. The Pace Generator process models the generation of pulses. The Update Timers process is used to simulate a global clock in SPIN, which enables us to specify the desired timing properties in LTL and verify them.

#### A. Detailed Model of the Pacemaker

In addition to the features described by the pacemaker models in [3, 4] our models have two extra processes viz. Accelerometer and Rate Controller. This enables us to model advanced features like rate controlled pacing and hysteresis. Next we describe all the processes and the behaviors of the cardiac pacemaker that they capture.

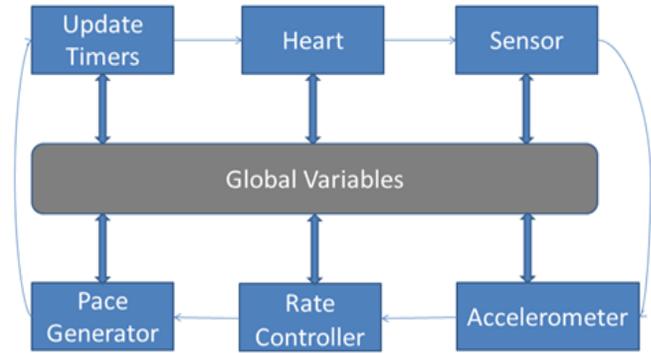

Figure 2. Detailed model of pacemaker with various components

*Update Timers* – This process maintains a global clock which is incremented in every round. The clock is used by other processes to express timing requirements of the system. It is reset after one full AV (Atria and Ventricles) cycle, which constitutes one pacing of both chambers of heart. Also some of the global variables which capture the pulses and senses are reset along with the various timestamps which capture the time of pulses and senses.

*Heart* – The heart process models the environment in which the pacemaker is run, we model the following four different behaviors of the heart, where, NR is Normal Rate, A is Atria, V is Ventricles and AVD is AV Delay:

Normal – Wait NR, Pace A, Wait AVD, Pace V, Repeat
Miss Ventricle Pace – Wait NR, Pace A, Wait AVD, Skip, Repeat
Dead – Wait NR, Skip, Wait AVD, Skip, Repeat
Non Deterministic – Wait NR, May Pace A, Wait AVD, May Pace V, Repeat

*Sensor* – This process captures the paces from the Heart and the Pace Generator and acts as a pulse sensor. It also records the time of the various pulses which is used to verify the refractory period property.

*Accelerometer* – This is another sensor in the pacemaker system which detects the motion of the human body. Based on the raw acceleration data there are 5 different activity thresholds mentioned in the informal specification [1]. We use these thresholds and set the response factor (RF) which is in turn used by the Rate Controller to set the appropriate rate of pacing.

*Rate Controller* – This process controls the rate at which the artificial pulses are delivered by the Pace Generator. In an event of increased body activity such as during exercise, the Accelerometer would detect the motion and set up the appropriate RF, which is used by the Rate Controller to determine the pacing rate. This rate is used by the Pace Generator to wait for the given time before delivering the next pulse.

*Pace Generator* – This is the most important component of the pacemaker system. Pace Generator is responsible for generating the pulses according to the mode of operation of the device. In total there are 18 different modes, these modes specify which portions of the heart are sensed and/or paced, how the paces are delivered and the whether they are rate controlled. Each mode can be written as WXYZ where,

W – Specifies which chambers of heart are paced and can take values A (Atria), V (Ventricles) and D (both).
X – Specifies which chambers of heart are sensed and can take values A, V, D and O (not specified).
Y – Specifies how the senses are handled and can take values I (Inhibited), T (Triggered), D (Tracked) and O.
Z – Specifies if the pacing is rate controlled or not (value R for Rate Controlled and unspecified otherwise).

The normal operating modes of the pacemaker are described in [2, 3]. For our PROMELA based pacemaker model we illustrate only the following modes which are new and not captured by previous models.

*Rate Controlled Pacing* – In rate controlled pacing the rate at which the paces are delivered is varied according to the activity level of the human body which is detected using the Accelerometer. There are 8 rate controlled modes viz. VOOR, AOOR, DOOR, VVIR, AAIR, DDIR, VDDR and DDDR. As mentioned earlier the Rate Controller sets the rate for pacing based on the RF. How this RF can be used to vary the rate is illustrated in Fig. 3 for the VDDR mode. When the Pace Generator is in VDD mode we wait for duration of Min Time before going from state 1 to 2, now this interval is based on the rate decided by the Rate Controller. So, before going to state 2 we wait for Min Time + RF*Increment, where Increment represents the minimum increment allowed in the change of rate. The RF values are set so that the following relation holds: Min Time + max (RF)*Increment < Max Time. This ensures that the Rate Controller does not change the rate beyond what is allowed by the Max Time. Max and Min Time are calculated based on the LRL and URL parameters given in [1] and RF is the response factor as described in [13]. Similarly for other modes we can include rate controlled pacing by changing the time between each paced pulses.

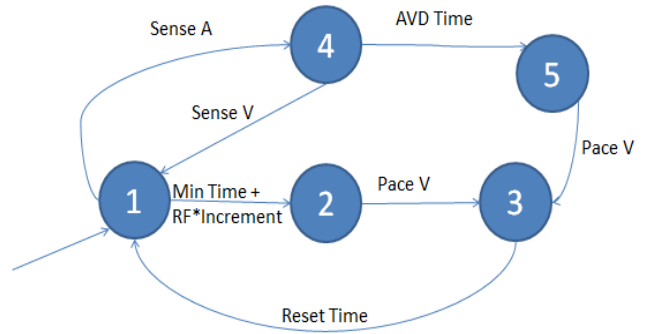

Figure 3.     VDDR mode

*Hysteresis Pacing* – This feature is valid only for inhibiting and tracking modes (with or without rate controlled pacing). When Hysteresis mode is turned on after every sense there will be a longer time period before the next pace. This will make sure if there is another sense then it will inhibit the pending pace. This longer time period to wait for the next pace can again be set using the RF to the maximum value. This will ensure that after every sense there is a longer time period of waiting. This is illustrated in Fig. 4 for the VDDR mode, at state 4 we wait for a longer period which is determined by AVD Time + RF*Increment before we go to the state 5. The rest of the diagram is the same as Fig. 3. Similarly for other modes we can use the RF to set the time period for hysteresis pacing.

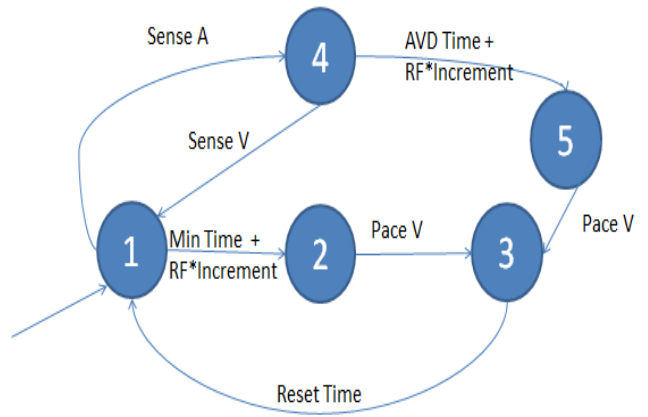

Figure 4.     VDDR mode with hysteresis

### B. LTL Properties

In the previous section we saw how we can model various modes of the pacemaker in SPIN using PROMELA. The desirable properties of the system can be written in

LTL, which can be used by SPIN to verify the PROMELA model. In the following section we describe the various LTL properties [13] which we verified in our model. Most of these properties represent timing requirements which are captured from the specification.

*Deadlock* – This property captures the fact that there is no deadlock in the system. The property of deadlock is actually specified by default in SPIN when it checks for invalid end states of the model. This ensures that all the processes are indeed deadlock free.

*Pace Limit* – There is an Upper Rate Limit (URL) and a Lower Rate Limit (LRL) specified in the requirements which means that the rate of pacing at any time should be between these two values. In order to ensure this we construct two LTL formulas, LRLURLA.ltl and LRLURLV.ltl, where LRLURLA.ltl can be written as, **G** (Rate of pacing A < URL && Rate of pacing A > LRL) this property captures the Pace Limit for Atria. Similarly we can construct the LTL formula for Ventricles.

*AV Delay* – This property represents the fact that between every A and V pulse the delay is always less than a fixed time period. Hence we can construct an LTL formula AVD.ltl as, **G** (AV_Delay < Fixed_AVD)

*Refractory Period* – The time taken between a sense and a pace in that chamber is called the Refractory Period; this property ensures that there is always some time delay between subsequent paced events in the Heart. There are three such properties which we capture in our model: ARP.ltl, VRP.ltl and PVARP.ltl, where ARP.ltl can be a LTL formula of the following kind: **G** ((Last_PacedA – Last_SensedA) > ARP). VRP.ltl and ARP.ltl can be defined similarly.

*Inhibiting Property* – This property is valid for inhibiting pacing modes and checks if the pending paces are inhibited in the presence of a sense in that chamber. These are represented by two LTL formulae AAI.ltl and VVI.ltl. Where AAI.ltl can be constructed as **G** (sense A -> not pace A).

*Triggering Property* – This property is valid for the triggered pacing modes and checks if the paces are triggered in the chamber whenever a sense is detected in that chamber. These are represented by two LTL formulae AAT.ltl and VVT.ltl. Where AAT.ltl can be written as **G** (sense A -> pace A).

*Tracked Property* – This property is valid for the tracked pacing modes and checks if the tracked pace V is delivered after a sense A and inhibited if there is a sense V before that. It is represented by the LTL formula XDD.ltl and can be written as **G** (sense A - >F (pace V && AV_Delay < Fixed_AVD)).

The properties described above are the seven basic properties of the pacemaker system which capture the timing constraints and the functions of various modes. Out of these seven the first four properties, Deadlock, Pace Limit, AV Delay and Refractory Period are generic properties of the system, while the last three are only valid for specific kinds of modes. In addition to these seven properties we also describe in our report [13], three advanced properties which are valid for rate controlled and hysteresis pacing modes. Thus we have a total of 10 desired properties of the system represented by 18 LTL formulas. We use the SPIN model checker to verify these properties for our PROMELA model; the results are discussed in the next section.

## IV. VERIFICATION OF PACEMAKER MODEL

We use SPIN to verify the various properties described in the previous section. Table I collects the results of verifying these properties in our model. A tick mark indicates that we were able to verify the property while a blank indicates that the property does not make sense for that particular mode. As is clear from Table I, we were able to verify the generic properties and also the mode specific properties in our model. Since the focus here is to validate the timing constraints of the Pace Generator we ignore particular heart behavior if it does not make sense for a particular mode (e.g. AOO mode, Dead Heart and Pace Limit for V). In section VI D we describe how to derive consistent operating modes of the pacemaker based on various heart conditions.

TABLE I.
VERIFICATION OF LTL PROPERTIES OF PACEMAKER PROMELA MODEL

| LTL Property | VOO | AOO | DOO | VVI | AAI | DDI | VVT | AAT | VDD | DDD |
|---|---|---|---|---|---|---|---|---|---|---|
| **Deadlock** | √ | √ | √ | √ | √ | √ | √ | √ | √ | √ |
| **Pace Limit** | √ | √ | √ | √ | √ | √ | √ | √ | √ | √ |
| **AV Delay** | | | √ | | | √ | | | √ | √ |
| **Refractory Period** | √ | √ | √ | √ | √ | √ | √ | √ | √ | √ |
| **Inhibiting** | | | | √ | √ | √ | | | | |
| **Triggering** | | | | | | | √ | √ | | |
| **Tracking** | | | | | | | | | √ | √ |

## V. C CODE GENERATION

The syntax of PROMELA [10] is very close to that of C but with some distinct features like non deterministic branching and communication using channels. This has encouraged the development of many C to PROMELA translators [11, 12] but only a few systems [5] exist for PROMELA to C code generation. SPIN internally translates PROMELA to C before execution and simulation of a model. The system in [5] uses the code from State Transition Matrix built internally by SPIN (pan.m and pan.t) to generate a C implementation which simulates the behavior of the PROMELA model. For our purposes this is not appropriate, as the use of an explicit State Transition Matrix would generate a very large amount of C code. Moreover the target pacemaker hardware platform is a PIC based microcontroller which has many restrictions on the kind of C that can be compiled and run on it (Small Device C compiler). In order to avoid building of an explicit State Transition Matrix we describe a refinement based translation from PROMELA to C.

Fig. 5 shows two kinds of rules we use to refine a PROMELA model in an implementation based on C. Due to lack of space we list the rules only for some of the more useful PROMELA constructs. The premise of a rule matches a given PROMELA construct which is replaced by the corresponding C code at the bottom. The rules are grouped together based on whether they match a data structure or a control structure.

Rules d1 to d8 are based on data refinement, they translate from PROMELA data structures to corresponding ones in C. Most of the rules are straight forward mapping of the appropriate data type in C. Rule d5 shows how to refine a 'mtype' as a series of '#define' declarations. This is done to reflect the semantics of the 'mtype' data type in PROMELA. Rule d8 refines a PROMELA channel 'chan' to a queue which is represented as a buffered array in C of corresponding size. For control refinement we have rules c1 to c8 which translate the appropriate control flow constructs from PROMELA to C. Rules c3 and c4 refine the send and receive operations on channels with enqueue and dequeue functions which act on the C array refined via rule d8. Non-deterministic choice in PROMELA is treated as under specification (or external input) of the model and refined as a stub function which is called to generate the choice deterministically as shown in rule c7. For now, to mimic the random simulation of the PROMELA model we choose a value randomly and use a switch case to take the corresponding branch. This enables us to keep the same behavior as the model if we were to validate this C code. Later this stub function is supposed to be implemented separately and added to the generated code. The interleaving between various PROMELA processes are refined using POSIX threads. We translate each process as a C function and then create a thread to execute that function as shown in rules c6 and c8. This set of 16 refinement rules is sufficient to translate most of the common used constructs in PROMELA to C code. These rules are similar in spirit to the ones described in [14] for refinement from B specifications to C.

Now we show that the refinement scheme of Fig. 5 preserves the LTL properties from PROMELA to the generated C code. We use the interpretation of a PROMELA model as a state transition system. Each process in PROMELA can be represented as a finite state automaton of the form (S, s0, L, T, F) where,

- S is set of states which represents all possible points of control within a process.
- T is the transition relation which defines the flow of control within a process.
- L is the set of Labels, which relate each transition to an update statement.
- F is the set of final states
- s0 is the initial state

We proceed first by showing that the finite state automaton represented by each of the generated C functions is equivalent to the corresponding one for the PROMELA processes. Then we argue that the global execution semantics of SPIN is an asynchronous composition of these automata while in C it is just a scheduled composition and thus is a subset. This ensures that if a LTL property is verified on the PROMELA model it will still be valid for the C implementation generated by following the refinement rules of Fig. 5. Let the generated C function be represented by an automaton D' = (S', s0', L', T', F'). And the corresponding state transition of the PROMELA process be D = (S, s0, L, T, F). Now based on Fig. 5 we see that the control refinement rules c1, c2, c5 and c6 do not change the control flow between PROMELA and C, thus the generated function has the same number of control locations as the process. In other words, we have S' = S and s0' = s0. Now in case of read and write statements (for a channel) the only difference is that while reading or writing to channel we have a for-loop in the C code (rules c3 and c4), which was just a single statement in the PROMELA process. This does not lead to a new state but more transitions to the same state in the D'. In rule c7 we are choosing any one of the branches based on a function call which is similar to a non-deterministic choice in PROMELA.

So, if we have a transition from s to t in D given by, t = T(s,L) where Label L is an executable statement, and in D' for each execution of the for-loop we have s'[i] = T'(s'[i-1], L'). Then each of this s'[i] is indistinguishable from t. Thus all these states can be represented by a single state t'=t. For all the other states in D and D' we have T(s,L) = T'(s',L'). Hence D and D' are equivalent, in other words, the finite state automaton represented by the PROMELA process and the generated C code is equivalent. A given PROMELA model consists of several such processes, the execution semantics of SPIN is such that the resulting global transition system is an asynchronous composition of the processes. So if D1,D2,D3 … represent the finite state automata of the processes 1,2,3 … in SPIN, then the execution semantics of the model can be captured by Asynchronous_Composition(D1,D2,D3…) . Similarly for the generated C code if D'1, D'2, D'3 … are the finite state automata of the functions generated by translating processes 1,2,3… using the refinement scheme in Fig.5, then the execution semantics of the C program can be thought of as Scheduled_Composition(D'1,D'2,D'3…) (scheduled by the POSIX Thread scheduler). We already showed that D and D' are equivalent for all processes and corresponding functions. And POSIX scheduling [16] based on SCHED_FIFO (first-in-first-out) or SCHED_RR (round-robin based) Composition is a subset of Asynchronous Composition; hence the global transition system of the C program is a subset of the global transitions system of the PROMELA model. In the PROMELA model we have verified the LTL properties (listed in section IV), so those properties are also valid for the generated C program.

**Data Refinement**

```
skip       bool       byte         name [ const ] = expr
─────d1  ─────d2  ─────d3   ──────────────────────────────d4
  1         bit       uchar        name [ const ] = {expr, expr ...};

           mtype = { x1,x2,x3, ... xn }                    mtype var
   ──────────────────────────────────────d5     ──────────────────────d6
   #define x1 n; #define x2 n-1; ...  #define xn 1;        int var

      Typedef t { decl_list }           chan name = [ const ] of { t1,t2, ...}
   ──────────────────────────d7     ─────────────────────────────────────────d8
      struct t { decl_list };          struct channel_i {t1 var1,t2 var2,...};
                                        channel_i name [ const ];
```

**Control Refinement**

```
if :: sequence [ :: sequence ]* fi       do :: sequence [ :: sequence ]* od
──────────────────────────────────c1    ──────────────────────────────────c2
   { sequence; [sequence]* }             while(1) { sequence ; [sequence]* }

   name ! expr_1,expr_2,...,expr_n         name ? var_1,var_2,..., var_n
──────────────────────────────────c3    ──────────────────────────────────c4
for (int i = 1; i < n i++)              for (int i = 0; i < n; i++)
{ enqueue(name, expr_i); }              { var_i = dequeue(name); }

  ::expr1 -> expr2                        name (args) { sequence }
──────────────────────c5                ──────────────────────────────c6
if (expr1) expr2;                       void name (args) { sequence }

  ::expr_1                              init { run process1();
  ::expr_2 ...                                 run process2();...}
──────────────────────────c7            ──────────────────────────────────c8
switch (stub_func_i()){                 void main() {
case 1 : expr_1 ;                       pthread_t thread1, thread2, ...;
case 2 : expr_2 ;   ...}                pthread_create(&thread1,NULL,process1,args);
int stub_func_i()                       pthread_create(&thread2,NULL,process1,args);
{ return 1+rand()%(n-1);}               ...
                                        pthread_join(thread1, NULL);
                                        pthread_join(thread2, NULL);
                                        ... }
```

Figure 5.    Translation scheme used in refinement from PROMELA to C

## A. Validation of C Code

In the previous section we described how our translation scheme preserves LTL properties from a PROMELA model to generated C code. We can check that it is indeed the case by doing validation of the generated C code for the pacemaker system using SPIN. From version 4.0 onwards SPIN allows the use of embedded C code as part of the PROMELA model. This embedded code is treated syntactically and executed as a single step by SPIN. But this can still be used to validate parts of implementation directly as shown in [6]. For the pacemaker model we use the generated implementation of the processes one at the time and use them to validate the system against the PROMELA model. For example consider the Pace Generator process in our model from section II. This process will generate a function with the name of Pace Generator () in the implementation. Let us first see how we can try to validate this function using SPIN against the PROMELA model.

Since the PROMELA model has a process of the form proctype Pace Generator (), we remove all the code inside the process and add the code from the generated C implementation using the c_code { } construct in SPIN. This embedded C code is now part of the PROMELA model and since the rest of the processes are still the same we can validate the LTL properties which would use this C code as part of a process. There is still one important detail missing, SPIN cannot access the C variables defined inside the embedded C code. So, we add extra code at the end of embedded C code from Pace Generator () function which will update the corresponding PROMELA variables from the values changed through the C function. Using the now.variable_name construct we can update PROMELA variables at the end of function. Now when we try to verify the properties in SPIN the values of the corresponding variables are updated through the actual implementation code embedded inside a process in SPIN. Hence we have validated that the implementation of Pace Generator () indeed satisfies all the desired properties. Next we do the same for the other functions from the implementation. Each time we remove a single process from PROMELA model and replace it with the corresponding function from the generated implementation. This is illustrated in the following schematic.

```
For all functions in the C Code, do
-Replace the code from the corresponding
 process in PROMELA with the C code from
 the function.
-Verify LTL properties using SPIN
```

With this we end the discussion of our approach for doing end to end verification and validation with SPIN. In the next section we evaluate our approach.

## VI. EVALUATION AND RESULTS

Our full system for the cardiac pacemaker [13] consists of three different models of pacemaker covering all 18 operating modes and a set of 10 desirable LTL properties. We are able to model more modes and properties when compared to [3, 4] thus providing a more comprehensive solution. The incremental model of development described in [3] was useful for creating different models – sequential, concurrent and distributed for validating different aspects of the system. The refinement based translation of section V enabled us to generate an implementation as well. Each of our models is increasingly more complex and expresses more behavior. This is clear from Table II which shows the number of states searched for proving deadlock freeness in the model by SPIN.

TABLE II. NUMBER OF STATES EXPLORED BY SPIN

| Model  | Sequential | Concurrent | Distributed |
|--------|------------|------------|-------------|
| States | 392,716    | 35,684,919 | 125,373,000 |

In the following discussion the pacemaker system refers to the full system including all the models, properties and implementation code. Since our goal was to show that end to end verification and validation with SPIN is feasible we describe the evaluation based on the different phases of development and the bugs found or defects identified in that phase. For our purpose we divide the phases of formal development in the following - Modeling (informal specification to formal PROMELA model), Verification (proving LTL properties on PROMELA model) and Validation (C code generation using refinement rules and validation against PROMELA model).

### A. Formal Modeling

During the modeling phase we were able to identify several ambiguities in the specification [1] for the pacemaker system. While building the Rate Controller and the Accelerometer models we identified inconsistencies in specification on the use of the parameter Response Factor (RF) for rate controller pacing. Also the description about hysteresis mode was not clear in the original specification. Based on our models and supplementary information about the pacemaker system [3-5] we were able to resolve these discrepancies. We eventually were able to use the parameter RF for rate controlled as well as hysteresis pacing as shown in Fig.3 and Fig.4. We designed the sequential model to represent advanced modes (like rate controlled pacing and hysteresis pacing) and the concurrent (as well as distributed) model was created to detect race conditions and synchronization issues. Our aim was to develop models of increasing flexibility and expressiveness; to that end we were successful as depicted in Table II.

### B. Verification

The verification of pacemaker models in SPIN was used to prove desired LTL properties. As discussed in section III B and IV we verified 10 different properties which were based on timing constraints derived from the specification. During the verification of different models we identified a defect in the formulation of the AV Delay property. Recall from section III B that this property was stated as the LTL formula G (AV_Delay < Fixed_AVD). While verifying this property on the distributed model we identified a race condition between the Pace Generator and the Rate Controller processes which would lead to the violation of the AV Delay property. In order to fix this we added synchronization between the two processes for the AV Delay. We also added another LTL property Distributed AV Delay [13] which checks for LTL formula G (AV_Delay < Fixed_AVD) within each process in the distributed model. This ensures that the AV Delay property is valid for the whole system. Similarly in the concurrent model several race conditions were fixed by ordering the sequence of read and writes to the global variables [13]. Had we not considered multiple models of the pacemaker we would have missed some of these constraints.

### C. Validation

The generated C code was also validated for the desired LTL properties. Since our refinement scheme (Fig. 5) preserves the LTL properties we were able to validate the implementation successfully. However, our pacemaker models also include the simulation of the environment in form of the Heart process. So when we decided to check the system by using execution traces of timed pulses instead of the Heart process we were able to identify a few bugs related to refractory period by using the validation technique described in section V A. In a real execution trace the Sensor takes a finite amount of time to detect a pulse; during that time interval there should be no other activity of the pacemaker system. This aspect of the pacemaker was not modeled by the environment simulation which we did in the Heart Process. In order to fix this bug we added appropriate VRP, ARP and PVRP parameters in the system which capture the refractory periods in the pacemaker after a, V pulse, A pulse and between A and V pulses, respectively.

We were able to use SPIN as a formal tool during all the stages of development of the pacemaker system. It was helpful to find various bugs at every stage thus making this an end to end process for formal development of the system. However, formal methods are useful not only to find defects but also to understand and gain insight about the underlying system itself. Provided that the model is expressive enough we should be able to use the pacemaker model and derive operating parameters for various modes of the pacemaker. In the next section we describe the results of such an approach using SPIN which we applied on our pacemaker system to generate consistent operating modes.

### D. Deriving Consistent Operating Modes for Pacemaker

Our pacemaker system also includes a simulation of environment in form of the Heart Process. In section III A, we described various heart conditions that can be simulated by the model. Out of the all operating modes of the pacemaker only a few make sense for a particular heart condition. Since our model incorporates both the operation modes and heart conditions we can derive the consistent operating modes by constructing a LTL formula of the form G (Heart_Condition = missA && Operating_Mode = all -> AV Delay). This formula will force SPIN to check through all the modes for which the AV Delay property is valid. And if the property is not valid the counter example will include the assignment of Operating_Mode for which this property is not valid. While doing this we assume that there are no bugs in the model itself.

Based on this technique we were able to derive the operating modes for 3 different heart conditions as shown in Table III. By consulting the specification [1] we find that it is indeed the case that for these particular heart ailments the appropriate pacemaker operating modes are the same as derived by our system. Thus we have shown that our PROMELA model and thus the generated C code are expressive enough to be useful in deriving consistent operating modes for the pacemaker.

TABLE III. CONSISTENT OPERATING MODES FOR HEART CONDITIONS

| Heart Condition | Simulated Parameter | Derived Pacemaker Modes |
|---|---|---|
| Atrial Bradycardia | missA | AOO, AAI, AAT |
| Ventricular Bradycardia | missV | VOO, VVI, VVT |
| Atrioventricular Nodal | dead | DOO, DDI, VDD, DDD |

### E. External Validity and Refinement Checking

The description of our study so far is based on the pacemaker challenge problem. In order to show that our results can generalize to other problem domains beyond the cardiac pacemaker system we conducted some experiments for external validity. The key aspect of our approach is the refinement based translation from PROMELA to C which preserves the LTL properties which are verified for the PROMELA model. We picked a set of existing case studies which described safety and liveness property verification for a given PROMELA model. Then we refined the PROMELA model using the technique described in section V to generate an implementation in C. We then validate the same set of safety and liveness properties by using embedded C code feature in SPIN. Not only does this enable us to check if the refinement rules we presented are general enough to be used beyond the cardiac pacemaker system but it also shows that we can use SPIN to do refinement checking between the formal model and generated implementation. Table IV lists the case studies we used for refinement checking. These case studies are described in [18]; the modeling column shows the different aspects of the system that can be modeled in SPIN. This set of case studies has examples from various kinds of system that are modeled using PROMELA.

TABLE IV. REFINEMENT CHECKING

| Case Study | Modeling |
| --- | --- |
| Channels as Data Structures | Data Refinement |
| Eight Queen's Problem | Nondeterministic Algorithm |
| Rate Monotonic Scheduling | Real Time System |
| Fischer's Algorithm | Critical Section |
| Chandy-Lamport Algorithm | Distributed System |

We took the source for the PROMELA model and property verified on that model for each of these cases from the website of [20]. Then we used our technique for refinement to generate an implementation in C, which is compared with the corresponding PROMELA model and validated using SPIN (as described in section V A). We were able to successfully refine and validate these PROMELA models based on our 16 refinement rules described in section V. We do not claim that our method can be used for all kinds of critical systems but these examples which include real time and distributed systems show that the approach is generic enough to be of practical use.

We note that there are other threats to validity of this approach. In particular, it may not be always possible to express the desired properties of the system in LTL. In that case our method is limited to end to end validation of temporal properties expressed in LTL. The refinement of PROMELA model to C code is dependent on the availability of POSIX runtime on the target hardware. This may not always be possible. The generated code may not be used as is in that case, but existing techniques from code refinement [19] can be used to generate an implementation which is executable on the target hardware. Programs which use pointers to directly manipulate the memory cannot be generated from this approach, but the use of pointers for critical systems is not encouraged and better avoided. During formal modeling phase it is often required to simulate the environment to drive the model (e.g. Heart in case of pacemaker), so the generated C code includes the simulation of the environment which has to be replaced by external input. This implementation may lead to some errors when used with an external input which was not simulated.

The way we handled this problem was to use traces from external input along with the generated implementation from the model and validate it in SPIN. It may not be possible to do so in every system. However, modeling part of the environment is often desired as it helped us to derive consistent operating modes for the pacemaker.

VII. RELATED WORK

The cardiac pacemaker challenge has been known since 2007 and there have been three published papers which describe the attempts to solve it. The first is due to H. D. Macedo, P. G. Larsen and J. Fitzgerald [3], in this paper they show it is possible to evolve the model of the pacemaker system from a sequential to concurrent and finally to a distributed model. In various models different aspects of the system are verified and validated. Though our approach is inspired by them, it is different in two important ways: Firstly we specify and verify more properties in our model and secondly their approach uses VDM while we use SPIN and due to which we are able to verify the implementation of C code as well (which is supported by SPIN). In [2] A. O. Gomes and M. V. M. Oliveira show how it is possible to formally specify the cardiac pacemaker using Z, they fail to capture advanced features like rate controlled pacing and hysteresis which we are able to model in our system. The work which comes closest to ours is [4] where L. A. Tuan, M. C. Zheng and Q. T. Tho use CSP to model the system and PAT to verify the desired properties of the pacemaker. We are also able to model and verify all the properties they describe and in addition we can also validate the C Code against the same properties.

Similar refinement technique for persevering LTL properties has been developed for the Z specification language recently [17]. This makes it possible to use it to develop an implementation for the formal model described in [2]. We believe that it will correspond closely with the method described in this paper for end to end validation. Formal code refinement based on predicate transformer calculus as described in [19] is complementary to our approach. The refinement calculus of [19] cannot be directly applied to the PROMELA model but can be used for further refinement of the generated C code. Generating implementation from a formal model has been supported in other systems like Event-B [21], but we do not know of any other similar method proposed for PROMELA and SPIN. The advantage of using SPIN is that it can verify safety as well as liveness properties while Event-B only supports invariance properties and has to be extended in order to express handle liveness [22].

The use of SPIN for formal modeling of pacemaker was partially driven by the flexibility and comprehensiveness of the tool. SPIN is a well-known model checking tool which has been under constant development and improvement. This paper describes the first such attempt that we know of using SPIN to validate a cardiac pacemaker system. SPIN has successfully been used to verify important mission critical software like the NASA Mars Rover [7]. The pacemaker model we present in this paper is modeled as PROMELA specification in SPIN; in addition to PROMELA code, SPIN (version 4 onwards) also allows limited use of embedded C as part of the model. We exploit this feature to validate the C code we have refined from the PROMELA model. This allows us to validate the C code against the same LTL properties which were used to verify the PROMELA model, thus ensuring end to end verification of the whole system.

## VIII. CONCLUSIONS AND FUTURE WORK

In this paper we described an approach for end to end verification and validation by means of a solution to the pacemaker challenge, to the best of our knowledge it is the most complete and comprehensive attempt made so far on this problem. We built a sequential model of the pacemaker in PROMELA covering all the 18 operation modes and verified 10 desired properties of the system using SPIN. Then we generated an implementation from the formal PROMELA model by means of a refinement based translation scheme. We showed that the refinement scheme preserves the LTL properties of the original model. The generated C code was once again validated against the specification by a novel modular verification technique using embedded C code in SPIN. Thus we achieved end to end verification and validation of all of the operation modes in the pacemaker.

For future work some improvements can be made by adding more parameters including the non-timing related ones like Amplitude, Width etc. and support for other features of the Pace Generator like Noise, ATR and Diagnosis mode. More work needs to be done on the automation for the hardware implementation side as well, ideally one would want to specify the components in a high level modeling tool like SPIN and automatically generate code which can be compiled on Small Device C Compiler directly for the target hardware PIC microcontroller.

We have also demonstrated the feasibility of an approach for end to end verification by using only a model checker (SPIN). In future we would like to explore other applications of our approach in different problems of verification of reactive systems. Similar refinement techniques with other systems like B-Method have been shown to be quite successful as a formal development approach. Ours is a first step to bring some of the benefits to SPIN and PROMELA. Based on our experience with the cardiac pacemaker project we are confident that this approach can be applied to other systems with minor modifications. We have shown the applicability of this method for other similar systems using several existing cases studies on PROEMLA. In particular the refinement based translation scheme is a useful method for generating code which preserves the temporal properties verified for the model.

REFERENCES

[1] Pacemaker System Specification. In Boston Scientific 2007.
[2] A. O. Gomes and M. V. M. Oliveira. Formal Specification of a Cardiac Pacing System. In FM 2009.
[3] H. D. Macedo, P. G. Larsen and J. Fitzgerald. Incremental Development of a Distributed Real-Time Model of a Cardiac Pacing System using VDM. In FM 2008.
[4] L. A. Tuan, M. C. Zheng and Q. T. Tho. Modeling and Verification of Safety Critical Systems: A Case Study on Pacemaker. In ICSSIRI 2010.
[5] S. Loffler. From Specification to Implementation: A PROMELA to C Compiler. In Project Report Ecole Nationale Supérieure des Télécommunications.
[6] G. J. Holzmann. Logic Verification of ANSI-C code with SPIN. In SPIN 2000.
[7] G. J. Holzmann and R. Joshi. Model-Driven Software Verification. In SPIN 2004.
[8] C. M. Holloway. Why Engineers should consider Formal Methods. In DASC (Digital Avionics Systems Conference) 1997.
[9] M. Archer, C. Heitmeyer and E. Riccobene. Proving invariants of I/O automata with TAME. In ASE 2002.
[10] A. F. Donaldson and A. Miller. Automatic Symmetry Detection for Promela. In J. Autom. Reasoning 41(3-4): 251-293, 2008.
[11] C. Walton. Model checking agent dialogues. In Int. Workshop on Declarative Agent Languages and Technology 2004.
[12] K. Jiang and B. Jonsson. Using SPIN to Model Check Concurrent Algorithms, using a translation from C to Promela. In Second Swedish Workshop on Multi-Core Computing 2009.
[13] A. Sharma. Towards a Verified Cardiac Pacemaker. In Technical Report NUS November 2010.
[14] D. Bert, S. Boulme, M-L. Potet, A. Requet and L. Voisin. Adaptable Translator of B Specifications to Embedded C Programs. In FME 2003.
[15] Proposal for the Pacemaker Formal Methods Challenge hosted by SQRL. http://sqrl.mcmaster.ca/pacemaker.htm. Retrieved on 26$^{th}$ Sep 2011.
[16] Thread Scheduling. In IEEE POSIX 1003.1c standard.
[17] J.Derrick and G. Smith. Temporal-logic property preservation under Z refinement. In Formal Aspects of Computing (25$^{th}$ May 2011).
[18] M. Ben-Ari. Chapter 11 Case Studies. In Principles of Spin Model Checker 2008 edition.
[19] R-J. Back and J.v. Wright. Refinement Calculus – A Systematic Introduction 1998 edition.
[20] http://www.springer.com/978-1-84628-769-5.Source of PROMELA models for case studies. Retrieved on 27$^{th}$ Sep 2011.
[21] J-R. Abrial. In Modeling in Event-B: System and Software Engineering. Cambridge University Press 2010.
[22] O. Mosbahi and J. Jaray. Specification and Proof of Liveness Properties in B Event Systems. In ICSOFT 2007.